\documentclass[10pt]{article}
\usepackage{amsmath}
\usepackage{amsfonts}
\usepackage{amssymb}

\begin{document}

\author{E. G. Beltrametti$^\dagger$ and S. Bugajski$^\ddagger$\footnote
{Supported by the Polish Committee for Scientific 
Research (KBN), grant No 7 T11C 017 21.}\\
$^\dagger$Department of Physics, University of Genoa, and\\ 
INFN-Sezione di Genova\\ 
$^\ddagger$Institute of Physics, University of Silesia, and\\
Institute of Pure and Applied Computer Science, \\
Polish Academy of Sciences
 }
\date{}
\title{Correlations and entanglement in probability theory}
\maketitle

\begin{abstract}
We generalize the classical probability frame by adopting a wider family of
random variables that includes nondeterministic ones. The frame that emerges
is known to host a ''classical''
extension of quantum mechanics. We discuss the notion of probabilistic 
correlation and show that it includes two kinds of correlation: a classical one, 
which occurs for both deterministic and indeterministic observables, and a 
nonclassical one, which occurs only for indeterministic observables. The latter 
will be called probabilistic entanglement and represents 
a property of intrinsically random systems, not necessarily 
quantum. It appears possible to separate the two kinds of correlation and 
characterize them by numerical functions which satisfy a simple product rule.
\end{abstract}

\section{Introduction}

Though the notion of entanglement is an intensively discussed concept of quantum
mechanics, it is hard to find in the literature a precise definition 
of this concept (we refer to [4] for various characterizations of 
entanglement). Our attempt here is to develop a probabilistic formulation 
of the entanglement as a specific kind of correlation. 

The probabilistic 
framework we will adopt is summarized in Section 2: it rests on the usual 
Kolmogorovian formulation of probability theory but it allows a wider family 
of random variables in order to encompass indeterministic behaviours.

When we speak of correlations we always mean correlations among results
of a joint measurement of two or more observables at some state. In the 
classical deterministic case there is no joint observable other than 
the product joint observable, 
so that the occurrence of a correlation depends only upon 
properties that are coded in the state of the physical system. In particular, 
no correlation can occur when the state is pure, namely when we have to do 
with a maximal 
information about the preparation of the physical system. Things are 
different, as discussed in Section 3, when we deal with indeterministic 
observables: the choice of a joint observable is, in general, no longer unique 
and the occurrence of a correlation becomes a function of that choice. New 
kinds of correlations emerge and the notion of entanglement will appear 
related to the fact that a non-product joint obseravble can give rise
to a correlation even if the physical system is in a pure state.

\section{Probabilistic physical systems}

We provide here a probabilistic framework that appears general eonugh to 
host the description of physical objects showing an intrinsic random 
behaviour. As usual, 
the basic ingredients are the physical notions of states and observables, 
whose counterparts in the language of probability theory are the probability 
measures on the sample space and the random variables.  

The states of a probabilistic physical object are pragmatically identified with
the possible ways of preparing statistical ensembles of samples of the object 
under discussion. The set of all states carries the natural convex structure 
which mirrors the possibility of performing 
statistical mixing of ensembles. We assume
that the convex set of states can be identified with (is affinely isomorphic
to) the convex set $M_1^+(\Omega) $ of all probability
measures on a measurable space $(\Omega ;{\cal B}(\Omega))$, 
often called the {\it phase space} of the physical system (~${\cal B}(\Omega)$ 
stands for a Boolean algebra of measurable subsets of $\Omega$~).  
We assume further that every one-point
subset of $\Omega $ is measurable, so that $\{ \omega\} \in 
{\cal B} (\Omega) $ for every $\omega \in \Omega .$
The last assumption implies that all Dirac measures $\delta_{\omega },\:
\omega \in \Omega ,$ belong to $M_1^+(\Omega) ,$ so representing
particular states called {\it pure states}. Non-pure states are also called 
{\it mixed}. 

Notice that, due to the particular convex structure of $M_1^+(\Omega)$, 
a mixed state has the classical property of admitting 
a unique decomposition into pure states: this alludes to the 
fact that the family of observable properties of the physical system under 
discussion will be rich enough to separate the elements of $M_1^+(\Omega)$. 
There is no distinguished probability measure on $\Omega $;
all probability measures on $\Omega $ are considered as possible states of
the probabilistic (randomly behaving) physical system.

We restrict our attention to a special (nevertheless sufficiently large)
class of experiments, or {\it measurements} with random outcomes (see,
for instance, [6], [8]): any measurement on a probabilistic
physical system is performed on a state, so on an ensemble of identically
prepared samples of the physical system, and results in a probability
distribution on the space of the possible outcomes of single individual 
measurements. Equivalence classes of measurements are called 
{\it observables} (observable properties). 
Therefore, an abstract description of an observable is a map 
$$
A: M_1^+(\Omega) \rightarrow M_1^+(\Xi)
$$
which transforms states into probability distributions on the space $\Xi $
of the outcomes of the observable; the measurement of $A$ at the state 
$\mu\in M_1^+(\Omega) $ will produce the measure $A\mu \in M_1^+(\Xi) $
to be called the {\it outcome measure}. Clearly, the outcome space has to be
measurable: we will further 
assume that every one-point subset of $\Xi$ is measurable, 
namely, $\{ \xi\} \in {\cal B}(\Xi) $ for every $\xi \in \Xi .$

All observables we are going to consider react linearly on the mixing of
ensembles; thus, an observable should be an affine ({\it i.e}. convexity
preserving) map, explicitely: 
$A(\lambda \mu_1+(1-\lambda) \mu_2) =\lambda A\mu_1+(1-\lambda) A\mu_2$ 
for every $\mu _1,\mu _2\in M_1^{+}( \Omega) $ and every
real number $\lambda$, $0\leq \lambda \leq 1.$

It is natural (see [5]) to assume that every observable $A:M_1^{+}( \Omega)
\rightarrow M_1^{+}( \Xi) $ satisfies the condition
\begin{equation}
(A\mu)(X) =\int_{\Omega}(A\delta_{\omega })(X)\:\mu (d\omega)
\end{equation}
for every $X\in {\cal B}(\Xi) $ and every state $\mu \in M_1^{+}( \Omega).$ This
measurability condition expresses the specific property that an
observable is uniquely defined by the outcome measures it associates to 
the pure states.

\subsection{Deterministic and indeterministic observables}

An observable $A:M_1^{+}( \Omega) \rightarrow M_1^{+}( \Xi
) $ will be called {\it deterministic} if it assumes a well defined
value at every pure state; formally if it maps Dirac measures on 
$(\Omega;{\cal B}(\Omega))$ into Dirac measures on $(\Xi;{\cal B}(\Xi))$. 
In other words, the deterministic observables have no dispersion on pure 
states. The affine character of the observables implies that a deterministic 
observable can exhibit some randomness, or some dispersion, 
when it acts on a mixed state. However, this stochastic aspect is fully reducible to the
mixed character of the state. It is easy to see that deterministic
observables correspond just to $\Xi $-valued random variables of the
standard probability theory (see [5]).

Observables which are not deterministic exhibit an inherent
randomness: at some pure states they show dispersion, {\it i.e.}, they 
generate diffused probability measures on the outcome space.

The distinction between deterministic and indeterministic observables is
of a deep nature. The traditional notion of random variable mirrors our 
notion of deterministic observable: the classical probability theory 
which admits only deterministic
observables describes merely a lack of information in a
deterministic world. The concept of non-deterministic observable 
implies in general a non-trivial statistical dispersion of
possible results of an observation even in presence of a maximal information 
about the preparation of the physical system. The probabilistic framework 
here introduced, which includes the occurrence of indeterministic observables, 
belongs to the so called operational probability theory (see [2],[5]) and is
able to describe an irreducible randomness - either inherent or produced by
an uncontrolled outer influence. 

It is evident that typical 
quantum-mechanical observables have an indeterministic nature. It
has been shown elsewhere (see [2]) that the operational probabilistic 
framework here described can host an extension of quantum mechanics. Notice 
that if we call $\Omega$ the set of pure states of a quantum system, then the 
usual family of the quantum-mechanical observables is not rich enough to 
separate $M_1^+(\Omega)$: indeed we know that the usual quantum mixed states 
have infinitely many convex decompositions into pure states. In the extension 
of quantum mechanics alluded to, the quantum observables are embedded into 
the bigger family of the observables on $M_1^+(\Omega)$, which is actually 
rich enough to separate distinct convex combinations of pure states.

\subsection{Joint observables}

We have defined an observable without any specific characterization of 
its outcome space $\Xi$. When the outcome space takes the structure of 
the Cartesian product $\Xi =\Xi _1\times \Xi _2$, then the observable 
$A:M_1^{+}( \Omega) \rightarrow M_1^{+}( \Xi _1\times \Xi_2) $ defines two 
observables 
$$
A_i:=\Pi _i\circ A\;,\:\:\:i=1,2\;, 
$$
where $\Pi _i:M_1^{+}( \Xi _1\times \Xi _2) \rightarrow
M_1^{+}( \Xi _i) $ is the marginal projection, {\it i.e.}, 
$$
\Pi _1\nu ( X_1) :=\nu ( X_1\times \Xi _2) \:\:,\:\:\Pi
_2\nu ( X_2) :=\nu( \Xi _1\times X_2) \;, 
$$
for every $\nu \in M_1^{+}( \Xi _1\times \Xi _2) ,\:X_i\in 
{\cal B}( \Xi _i) ,$ $i=1,2$ .

The observable $A:M_1^{+}( \Omega) \rightarrow 
M_1^{+}(\Xi_1\times \Xi _2) $ is called a {\it joint observable} of $A_1$ 
and $A_2$. It should be stressed, however, that a pair of
observables $A_1:M_1^{+}( \Omega) \rightarrow M_1^{+}(\Xi_1) ,$ 
$A_2:M_1^{+}( \Omega) \rightarrow M_1^{+}( \Xi_2) $ admits, in general, 
many joint observables.

A particular joint observable for $A_1$ and $A_2$ is the product one,
denoted $A_1\boxtimes A_2$, and defined by 
\begin{equation}
A_1\boxtimes A_2\:\delta _{\omega }:=A_1\delta _{\omega }\boxtimes A_2\delta
_{\omega }
\end{equation}
for every $\omega \in \Omega $ (see [2],[5]), the symbol $\boxtimes$ in the 
r.h.s. standing for the usual product of measures (if $\nu_i\in M_1^+(\Xi_i),
\:i=1,2$, then $\nu_1\boxtimes \nu_2(X_1\times X_2):=\nu_1(X_1)\cdot \nu_2(X_2)$
for every $X_i\in{\cal B}(\Xi_i),\:i=1,2$).  The definition
is exhaustive thanks to the condition of measurability expressed by Eq.(1).
As the product observable $A_1\boxtimes A_2$ does 
exist for every pair $A_1,A_2$, any two observables pertaining
to a probabilistic physical system are comeasurable, what is commonly
considered as a feature characterizing ''classical'' theories. In this 
sense the present framework behaves as ''classical'' though only the 
deterministic observables have a counterpart in classical mechanics and in 
standard classical probability theory.

Any pair of deterministic observables has a unique joint observable, 
the product observable, in full agreement with the standard probability theory
(where the joint observable is usually called the ''random vector''), while
a joint observable of two indeterministic observables is non-unique.

Writing $J( A_1,A_2) $ to denote a joint observable 
of $A_1:M_1^+(\Omega)\to M_1^+(\Xi_1)$ and 
$A_2:M_1^+(\Omega)\to M_1^+(\Xi_2)$, the measure 
$J( A_1,A_2) \mu \in M_1^{+}(\Xi _1\times \Xi _2)$, 
{\it i.e}. the statistical result of the measurement of $J( A_1,A_2) $ at $\mu
\in M_1^{+}(\Omega) $, 
clearly determines the statistical results of measurements of
both $A_1$ and $A_2$ at the state $\mu .$ 
Thus, the joint observable $J(A_1,A_2) $ can 
be interpreted as a particular method for a simultaneous
measurement of $A_1$ and $A_2.$ Various joint observables would then
correspond to various ways of performing simultaneous measurements.

On the other hand, the fact that $A_1\mu $ and $A_2\mu $ do not determine 
uniquely $J( A_1,A_2) \mu $ suggests that a simultaneous measurement of 
$A_1 $ and $A_2$ could even be performed directly, without measuring any joint
observable of them. Indeed, one could measure $A_1$ on every second member of
the statistical ensemble $\mu ,$ and $A_2$ on the other members. Such a
measurement provides directly two probability measures $A_1\mu $ and $A_2\mu 
$ without any joint measure of them.

This means that the concept of a simultaneous
measurement of two or more observables does not coincide with the concept 
of a joint measurement. The
former is surely more general than the latter, but a
measuement of a joint observable is more ''informative'': it provides 
also the correlations that might emerge between, or among, the observables.

\section{Probabilistic correlations}

We assume the common-sense concept of correlation: two parameters are {\it 
correlated} whenever they are not independent. The notion of correlation  
can be understood in various ways, for instance in algebraic terms as a 
(non-trivial) relation, or as a fuzzy relation, but here we mean the 
probabilistic version of that concept; hence we speak about probabilistic 
(or stochastic) correlation and probabilistic (or stochastic) independence. 

Consider two parameters and let $\Xi_1,\Xi_2$ be the (measurable) sets 
of their possible values; given a measure 
$\nu \in M_1^{+}( \Xi _1\times \Xi _2)$ we say, according to the 
standard probability theory, that the two parameters are mutually independent 
relative to $\nu $ iff
\begin{equation}
\nu =\Pi _1\nu \boxtimes \Pi _2\nu,
\end{equation} 
explicitely, iff $\nu(X_1\times X_2)=(\Pi_1\nu)(X_1)\cdot (\Pi_2\nu)(X_2)$ 
for every $X_i\in M_1^+(\Xi_i),\:\:i=1,2$. 
If $\nu $ has not this product form, the two parameters will be considered 
{\it correlated} relative to $\nu $.

It is an immediate consequence of this definition the fact that the two 
parameters are independent at every Dirac measure $\delta_{(\xi _1,\xi _2)}$ 
because 
$$
\delta_{(\xi_1,\xi_2)}=\delta_{\xi_1}\boxtimes\delta_{\xi_2}\;,\mbox{ for 
every }\xi_1\in \Xi_1,\;\xi_2\in\Xi_2.
$$

Thus we see that two parameters can be correlated only relative to a measure 
$\nu \in M_1^+(\Xi_1\times \Xi_2)$ having the form of a mixture; notice 
however that the mixed nature of $\nu$ is not a sufficient condition to produce 
correlation (for instance, if $\Xi_1=\Xi_2$ we have independence relative 
to the symmetrized mixture $\frac {1}{4}(\delta_{(\xi_1,\xi_1)}+
\delta_{(\xi_1,\xi_2)}+\delta_{(\xi_2,\xi_1)}+\delta_{(\xi_2,\xi_2) })$).

Let us stress that our definition of correlation as the negation of 
independence does not overlap exactly with the traditional one assumed in 
standard probability theory, where a {\it correlation coefficient} is 
introduced which quantifies to what extent the joint distribution 
$\nu \in M_1^{+}(\Xi _1\times\Xi_2)$ deviates from a one concentrated on a 
line, so that $\nu$ is said to carry a correlation 
if such a coefficient does not vanish. If 
$\nu =\Pi_1\nu \boxtimes \Pi_2\nu,$ namely if the parameters  are independent 
relative to $\nu$, then the correlation
coefficient equals zero. This always happens if $\nu$ is a Dirac 
measure. However, the vanishing of the correlation coefficient does not 
imply that $\nu =\Pi_1\nu \boxtimes \Pi_2\nu $; simple examples can be found 
in textbooks (see, {\it e.g.}, [9], pp. 328 - 9). That means that the 
correlation coefficient provides only a coarse characterization of the 
concept of correlation adopted here.

Let us now come to the notion of independence and correlation of 
observables. A pair of observables, say $A_1:M_1^+ (\Omega) \rightarrow
M_1^+ (\Xi _1) ,$ $A_2:M_1^+ (\Omega) \rightarrow
M_1^+ ( \Xi _2) ,$ does not produce at a state $\mu \in
M_1^+ (\Omega) $ any well defined measure on the product of the
outcome sets $\Xi _1\times \Xi _2.$ Consequently, it makes no sense to speak 
of independence (or correlation) of two observables without reference to 
some joint observable: the concept of
independence should be relative to a particular joint observable, so to
a particular way the two observables are paired together. Thus, the notion of 
independence expressed by Eq.(3) brings us to the following definition:

\smallskip

Two observables $A_1:M_1^+(\Omega) \rightarrow 
M_1^{+}(\Xi_1) ,$ $A_2:M_1^+(\Omega) \rightarrow M_1^+(\Xi_2) $ are 
independent at the state $\mu\in M_1^+(\Omega)$ relative to the joint observable 
$J( A_1,A_2) : M_1^+(\Omega) \rightarrow M_1^+(\Xi_1\times \Xi_2) $ iff 
$J( A_1,A_2) \mu =A_1\mu \boxtimes A_2\mu $.

\smallskip 

Accordingly, we say that two observables $A_1$ and $A_2$ are correlated 
(mutually dependent) at the state $\mu$ relative to 
the joint observable $J( A_1,A_2)$ iff
\begin{equation}
J( A_1,A_2) \mu \neq A_1\mu \boxtimes A_2\mu . 
\end{equation}

Though familiar examples of correlated observables often refer 
to compound systems, it should be noticed that the above concept of correlation 
does not require any restriction about the nature of the physical system.

\subsection{Classical correlations and probabilistic entanglement}

Among the possible joint observables of 
$A_1$ and $A_2$ there is always the product $A_1\boxtimes A_2$ 
characterized by the property expressed by Eq.(2) which 
ensures the independence of any two observables $A_1,\; A_2$ 
at every pure state, relative to $A_1\boxtimes A_2$.
Clearly, for a nonpure state $\mu \in M_1^+( \Omega) ,$ the
measure $A_1\boxtimes A_2\:\mu$ has the form of a mixture and, as already noticed, 
we have, in general, 
\begin{equation}
A_1\boxtimes A_2\:\mu \neq A_1\mu \boxtimes A_2\mu .
\end{equation}
In view of
the independence of $A_1$ and $A_2$ at every pure state, the occurrence of
a correlation in $A_1\boxtimes A_2\:\mu $ has to be interpreted as due to 
properties coded in the mixed state $\mu $. Thus, we
can view the joint observable $A_1\boxtimes A_2$ as representing the
correlation-free way of pairing $A_1$ and $A_2$, because a correlation 
appearing in $A_1\boxtimes A_2\:\mu $ is generated by the state $\mu $, namely 
by the way the pure states are mixed together to form $\mu$.

In the standard classical case, where only deterministic observables come 
into play, the product $A_1\boxtimes A_2$ is known to be 
the unique joint observable of $A_1,A_2$: the correlation carried by 
$A_1\boxtimes A_2\:\mu$ thus becomes the only possible correlation. This is why 
this correlation, which depends only on properties coded in the mixed state $\mu$ 
of the physical system, will be called the {\it classical correlation}. 

Clearly, when we speak 
of the classical correlation carried by $A_1\boxtimes A_2\:\mu$ we are not 
committed to assume that $A_1$ and $A_2$ are deterministic: the classical 
correlation can appear also in a nonderministic framework. 

But when we deal with two indeterministic observables $A_1,A_2$, the product 
$A_1\boxtimes A_2$ is no longer the only possible joint observable. A joint 
observable $J(A_1,A_2)$ differing from $A_1\boxtimes A_2$ can produce at a 
state $\mu$ a probability distribution $J(A_1,A_2)\mu$ carrying an 
additional correlation, which will depend on the particular way the two 
observables are paired together to form $J(A_1,A_2)$. Thus, we expect that 
the correlation contained in the outcome measure $J(A_1,A_2)\mu$ can be 
described as coming from two mutually independent factors:
(i) the classical correlation inherent to the state $\mu$, and
(ii) the possible additional correlation, to be called {\it probabilistic 
entanglement}, introduced by the joint observable in question. 
Due to this second factor the 
total correlation contained in the distribution $J(A_1,A_2)\mu$ should be 
different for different joint observables of $A_1$ and $A_2$. 

Typical of the probabilistic entanglement, associated to a joint 
observable differing from $A_1\boxtimes A_2$, is the fact that it 
need not vanish at pure states. Since the classical correlation vanishes at 
the pure states, any correlation appearing at a pure state has to be 
recognized as a probabilistic entanglement.

The occurrence of correlations at pure states 
is a characteristic feature of the quantum behaviour: the "entangled states" 
of the quantum dictionary are indeed pure states (of a compound system) and 
correlations can appear at these states. Let us stress that familiar 
expressions like ''a state shows entanglement'' make sense only if a joint 
observable is implicitly referred to (compare 
with [7]) and, similarly, one could say that ''a joint observable shows
entanglement'' only if a state is implicitly referred to.

Notice that the familiar EPR, or Bell, correlation is indeed a correlation 
occurring at a pure state of a compound system (an "entangled" state) and 
relative to a joint observable (the tensor product of the two observables 
involved) which is not a product joint observable in the sense of Eq.(2). 
What we have called probabilistic entanglement encompasses the quantum idea of 
entanglement, casting it in a more general framework.

Summing up, given two observables 
$A_1:M_1^+(\Omega) \rightarrow M_1^+(\Xi_1)\;,\;
A_2:M_1^+(\Omega) \rightarrow M_1^+(\Xi_2),$ given a joint observable 
$J(A_1,A_2)\neq A_1\boxtimes A_2$, and given a state $\mu \in M_1^+(\Omega)$, 
we focus attention on three probability measures on $\Xi_1\times \Xi_2$: 
$$
A_1\mu \boxtimes A_2\mu\;,\:\:\:\:\:\:
A_1\boxtimes A_2\:\mu\;,\:\:\:\:\:\:
J(A_1,A_2)\mu\;.
$$
The measure $A_1\mu \boxtimes A_2\mu $ can be seen as the 
result of performing an independent measurement of $A_1$ and $A_2$ at 
$\mu $  and taking formally the product of the two obtained measures (a 
joint observable giving the result $A_1\mu \boxtimes A_2\mu $ when 
measured at $\mu$ need not exist). 
On the contrary, the measures $A_1\boxtimes A_2\:\mu $ and $J(A_1,A_2)\mu$ can 
be naturally seen as the result of measuring the product joint observable 
$A_1\boxtimes A_2$ and, respectively, a non-product joint observable 
$J(A_1,A_2)$ at $\mu$. Obviously, all three measures give back $A_1\mu$ and 
$A_2\mu$ as marginals. 

What we have called classical correlation mirrors the 
departure betwen the two measures $A_1\boxtimes A_2\:\mu$ and 
$A_1\mu\boxtimes A_2\mu$; what we have called probabilistic entanglement 
mirrors the departure between the two measures $J(A_1,A_2)\mu$ and 
$A_1\boxtimes A_2\:\mu$. The total correlation at a state $\mu$ relative 
to the (non-product) joint observable $J(A_1,A_2)$ mirrors the departure 
between $J(A_1,A_2)\mu$ and $A_1\mu\boxtimes A_2\mu$.
In the next Section the separation between classical correlation and 
probabilistic entanglement will be made clear by introducing their density 
functions.

\subsection{Correlation density functions}

We will be interested in comparing two measures, say 
$\tilde{\nu}, \nu$, on a measurable 
space having the form $\Xi_1\times \Xi_2$. The Radon-Nikodym, or density 
function, of $\tilde{\nu}$ relative to $\nu$ is a function 
$d\tilde{\nu}/d\nu:\Xi_1\times \Xi_2\to{\bf R}^+ $ such that
$$
\tilde{\nu}( X) =\int_{X}\frac{d\tilde{\nu}}{d\nu}\cdot\nu 
(d(\xi_1,\xi_2))\:\:\:\:\mbox{for every } X\in {\cal B}(\Xi_1\times \Xi_2). 
$$ 
This Radon-Nikodym derivative provides a 
complete description of the measure $\tilde{\nu}$ (given $\nu$), hence it also 
contains all informations on the correlations inherent to $\tilde{\nu}$.

Let us now come to the three measures $A_1\mu \boxtimes A_2\mu$, 
$A_1\boxtimes A_2\:\mu$, and $J(A_1,A_2)\mu$ discussed in the previous section,
and consider the three density functions from $\Xi_1
\times \Xi _2$ into ${\bf R}^+$ (provided they exist, see the Appendix):
\begin{equation}
\rho_c:=\frac{d(A_1\boxtimes A_2\:\mu) }{d(A_1\mu \boxtimes A_2\mu) },\:\:\:\:
\rho_e:=\frac{d(J(A_1,A_2)\mu) }{d(A_1\boxtimes A_2\:\mu) },\:\:\:\:
\rho_t:=\frac{d(J(A_1,A_2)\mu) }{d(A_1\mu \boxtimes A_2\mu) }
\end{equation}

The classical correlation is fully characterized by the density function 
$\rho_c$, which describes the departure between the measure 
$A_1\boxtimes A_2\:\mu$ associated to the product joint observable and the 
measure $A_1\mu\boxtimes A_2\mu$ in which $A_1,A_2$ 
behave as independent. Thus, we call $\rho_c$ the {\it classical-correlation 
function}. 

The departure between the measures $J(A_1,A_2)\mu$ and $A_1\boxtimes A_2\:\mu$ is 
fully described by the density function $\rho_e$ which thus 
characterizes the probabilistic entanglement of the observables $A_1,A_2$ 
at the state $\mu$ relative to the (non-product) joint observable 
$J(A_1,A_2)$. The density function $\rho_e$ will thus be called the 
{\it entanglement function}.

As already remarked, the total correlation inherent to 
$J( A_1,A_2) $ at $\mu $ emerges by comparing the measure $J( A_1,A_2) \mu $
with the measure $A_1\mu \boxtimes A_2\mu $ which carries no correlation. An
exhaustive description of this correlation is then provided by the 
density function $\rho_t$ which we will call the 
{\it total-correlation function}. 

On the other hand,
we have seen that the correlations shown by $J( A_1,A_2)\mu $ 
can be divided into two parts, characterized by the two
functions $\rho_c$ and $\rho_e$ introduced above. The three
correlation functions are connected together: indeed, the general theory of 
the Radon-Nikodym derivatives states that (see {\it e,g.} [1], Corollary 
2.9.4, or [3], Sect. 32) 
\begin{equation}
\rho_t=\rho_c\cdot\rho_e.
\end{equation}
This simple product formula provides an {\it a posteriori} support for the ideas expressed in the 
previous Section.

Notice that for a pure state $\mu =\delta_{\omega }$ we have: $A_1\boxtimes
A_2\:\delta_{\omega }=A_1\delta_{\omega }\boxtimes A_2\delta _{\omega },$
so that the classical-correlation function $\rho_c$ becomes the constant 
unit function and we get $\rho_t=\rho_e$.

The calculation of the correlation functions might be hard for a general case,
but concrete situations of interest often involve finite sets of values 
as well as pure states or finite mixtures, what makes the calculations 
handable.

Let the two sets $\Xi_1,\Xi_2$ be finite and let $\xi_1\in \Xi_1,\:
\xi_2\in\Xi_2$: if, for instance,  we are concerned 
with the entanglement function relative to the joint observable 
$J( A_1,A_2):M_1^+(\Omega) \rightarrow M_1^+(\Xi_1\times \Xi_2)$, then we can 
write
$$
(J(A_1,A_2)\mu)(X)=\sum_{(\xi_1,\xi_2)\in X}\rho_e\cdot 
(A_1\boxtimes A_2\:\mu) (\xi_1,\xi_2)
$$
for every $X\in \Xi_1\times \Xi_2$. This makes clear that the entanglement 
function $\rho_e$ will take the form
$$
\rho_e(\xi_1,\xi_2) =\frac{(J(A_1,A_2)\mu)(\xi_1,\xi_2)}
{(A_1\boxtimes A_2\:\mu)(\xi_1,\xi_2)}.
$$
If the state $\mu$ is a finite mixture of pure states, say 
$\mu =\sum_i \lambda_i\delta_{\omega_i}$, then the above numerator and 
denominator will take the corresponding form of finite 
sums, since $J(A_1,A_2)$ and $A_1\boxtimes A_2$ act affinely on $M_1^+(\Omega)$.

Let us finally remark that the density functions introduced before can be 
used to produce various numerical characterizations of the corresponding 
correlations.

\section{Appendix}

We collect here some technical results on the existence of the Radon-Nikodym 
derivatives used in Section 3.

\smallskip

\noindent LEMMA: If $\nu \in M_1^+(\Xi_1\times \Xi_2) $ is a discrete measure 
and $\nu_1:=\Pi_1\nu$, $\nu_2:=\Pi_2\nu$ 
are its marginals, then $\nu \ll \nu_1\boxtimes \nu _2$ ({\it i.e.}, 
$\nu $ is absolutely continuous w.r.t. $\nu _1\boxtimes \nu_2$).

\smallskip

\noindent {\bf Proof} Recall that the defining property of the product measure 
$\nu _1\boxtimes \nu_2$ is: 
$$
(\nu_1\boxtimes\nu_2)(X_1\times X_2) =\nu_1(X_1)\nu_2( X_2) 
$$
for every $X_1\in {\cal B}(\Xi_1),$  $X_2\in {\cal B}(\Xi_2)$. If 
$\nu(\xi_1,\xi_2)$ and $\nu(\eta_1, \eta_2)$ are both non zero 
($\xi_1,\eta_1\in \Xi_1,\:\:\xi_2,\eta_2\in\Xi_2$), then 
$\nu_1(\xi_1), \nu_1(\eta_1), \nu_2(\xi_2), \xi_2(\eta_2)$ are all non zero 
so that $(\nu_1\boxtimes \nu_2)$ is different from zero at all points of the 
set $\{\xi_1,\xi_2\}\times \{\eta_1,\eta_2\}$. This implies that 
$\nu(X)=0$ for some $X\in {\cal B}(\Xi_1\times \Xi_2)$ implies 
$(\nu_1\boxtimes \nu_2)(X)=0$, what is equivalent to 
$\nu\ll\nu_1\boxtimes\nu_2$ . $\hfill \Box$

\smallskip

Notice that the discreteness restriction is crucial as the following 
counterexample (suggested by P.J. Lahti and A. Dvurecenskij) shows. Take 
for $\Xi_1$ and $\Xi_2$ the real interval $[0,1]$, let $\nu_1$, $\nu_2$ be 
Lebesgue measures on $\Xi_1$, $\Xi_2$ respectively, and define 
$\nu\in M_1^+(\Xi_1\times \Xi_2)$ by $\nu(X_1\times X_2):=\nu_1(X_1\cap X_2)$ 
(where $X_2\subseteq \Xi_2$ is identified with the corresponding subset of 
$\Xi_1$) so that $\nu$ is concentrated at the main diagonal $\{(\xi,\xi)\:|\:
\xi\in [0,1]\}$ of the square $\Xi_1\times \Xi_2$. Clearly, $\nu_1$ and $\nu_2$ 
are marginals of $\nu$, however $\nu_1\boxtimes\nu_2$ is the Lebesgue measure on 
$\Xi_1\times\Xi_2$ that vanishes on the main diagonal so that $\nu$ is not 
absolutely continuous w.r.t. $\nu_1\boxtimes\nu_2$.

The property stated in this Lemma ensures the existence of the 
Radon-Nikodym derivative $d\nu / d(\nu_1\boxtimes\nu_2)$ 
which is (see [1], Theorem 2.5.5) $\nu_1\boxtimes\nu_2$-almost everywhere 
finite. This result can be applied to ensure the existence of 
the density functions $\rho_c$ and 
$\rho_t$ occurring in Eqs.(6), provided the measures $A_1\boxtimes A_2\:\mu$ and 
$J(A_1,A_2)\mu$ are discrete. 

Concerning the entanglement function $\rho_e$ we make use of the following 

\smallskip

\noindent COROLLARY: Let 
$J(A_1,A_2):M_1^+(\Omega) \rightarrow M_1^+(\Xi_1\times\Xi_2)$ be a 
discrete joint observable of $A_1$ and $A_2$. 
Then, for every $\mu \in M_1^+(\Omega)$, 
we have $J(A_1,A_2)\mu \ll A_1\boxtimes A_2\:\mu$

\smallskip

\noindent {\bf Proof} In view of Eq.(1)) we have that 
$(A_1\boxtimes A_2\:\mu)(X) =0$ implies $(A_1\boxtimes A_2\:\delta_{\omega })
(X) =0$, $\mu $-almost everywhere. Since the previous Lemma ensures that  
$$
J(A_1,A_2)\delta_{\omega }\ll A_1 \delta_{\omega}\boxtimes A_2\delta_{\omega}=
A_1\boxtimes A_2\:\delta_{\omega }
$$
we also have $(J(A_1,A_2)\delta_{\omega})(X) =0$, $\mu $-almost everywhere. 
Looking again at Eq.(1) we conclude that $(A_1\boxtimes A_2\:\mu)(X) =0$ 
implies $(J(A_1,A_2)\mu)(X)=0$ 
for every measurable subset $X\in {\cal B}(\Xi_1\times \Xi_2).$ $\hfill \Box $

\smallskip

This corollary ensures the existence of the entanglement function 
$\rho_e$ occurring in Eq.(6) for every discrete joint observable $J(A_1,A_2)$.

\section*{\bf References}

[1] H. Bauer, {\it Probability Theory and Elements of Measure
Theory}, Academic Press, London, 1981.

\smallskip

\noindent[2] E. G. Beltrametti, S. Bugajski, J. Phys. A: Math.Gen. 
{\bf 28} (1995)
3329 - 3343. S. Bugajski, Int. J. Theor. Phys. {\bf 35} (1996) 2229 - 2244.

\smallskip

\noindent[3] P. Billingsley, {\it Probability and Measure}, J.Wiley, New
York, 1979.

\smallskip

\noindent[4] D. Bru\ss: {\it Characterizing entanglement}, \texttt{%
arXiv:quant-ph/0110078 v1 12 Oct 2001}.

\smallskip

\noindent[5] S. Bugajski, Mathematica Slovaca {\bf 51} (2001) 321 - 342 
and 343 - 361.

\smallskip

\noindent[6] P. Busch, P. J. Lahti, P. Mittelstaedt, {\it The Quantum
Theory of Measurement}, Lecture Notes in Physics, 2nd Edition, 
Springer-Verlag, Berlin, 1996.

\smallskip

\noindent[7] A. Cabello, Phys. Rev. A {\bf 60} (1999) 877 - 880.

\smallskip

\noindent[8] A. S. Holevo, {\it Probabilistic and Statistical Aspects
of Quantum Theory}, Statistics and Probability {\bf 1}, North-Holland,
Amsterdam, 1982.

\smallskip

\noindent[9] P. E. Pfeiffer, {\it Concepts of Probability Theory}, 
McGraw-Hill, New York, 1965.

\end{document}